\documentclass[
 reprint,
bibnotes,
 amsmath,amssymb,
 aps,
prb,
]{revtex4-1}

\usepackage{graphicx}
\usepackage[usenames]{color}
\usepackage{dcolumn}
\usepackage{bm}

\begin{document}

\preprint{APS/123-QED}

\title{Relaxation of the resistive superconducting state in boron-doped diamond films}

\author{A. Kardakova}
\email{kardakova@rplab.ru}
\author{A. Shishkin}%
\affiliation{%
Physics Department, Moscow State Pedagogical University, Russia
}%

\author{A. Semenov}
\affiliation{
Physics Department, Moscow State Pedagogical University, Russia
}%
\affiliation{
Moscow Institute of Physics and Technology, Russia
}%

\author{G.N. Goltsman}
\email{goltsman@rplab.ru}
 \author{S. Ryabchun}%
 \affiliation{
Physics Department, Moscow State Pedagogical University, Russia
}%
\affiliation{
National Research University Higher School of Economics, Russia
}%

\author{T.M. Klapwijk}
\email{t.m.klapwijk@tudelft.nl}
 \affiliation{
Physics Department, Moscow State Pedagogical University, Russia
}%
\affiliation{Kavli Institute of Nanoscience,
Delft University of Technology, The Netherlands 
}%
\affiliation{Donostia International Physics Center, Donostia-San Sebastian, Spain 
}%
\author{J. Bousquet, D. Eon, B. Sac\'ep\'e, Th. Klein, and E. Bustarret}

\affiliation{Institut N\'eel, CNRS, Grenoble, France}
\affiliation{Universit\'e Grenoble Alpes, Grenoble, France}
\date{\today}

\begin{abstract}
We report a study of the relaxation time of the restoration of the resistive superconducting state in single crystalline boron-doped diamond using  amplitude-modulated absorption of (sub-) THz radiation (AMAR). The films grown on an insulating diamond  substrate have a low carrier density of about $2.5\times10^{21}~$cm$^{-3}$ and a critical temperature of about $2~$K.  By changing the modulation frequency we find a high-frequency rolloff which we associate with the characteristic time of energy relaxation between the electron and the phonon systems or the relaxation time for nonequilibrium superconductivity. Our main result is that the electron-phonon scattering time varies clearly as $T^{-2}$, over the accessible temperature range of  $1.7$ to $2.2$~K. In addition, we find, upon approaching the critical temperature $T_c$,  evidence for an increasing relaxation time on both sides of $T_c$.

\begin{description}
\item[Usage]

\item[PACS numbers]{74.70.Wz, 72.10.-d, 74.62.En, 74.25.N , 74.40.Gh}

\item[Structure]

\end{description}
\end{abstract}



\maketitle


\section{Introduction}

Electron-phonon (e-ph) scattering is well understood in clean bulk normal metals and is part of  standard textbooks on solid-state physics. 
Within that framework, it is also embedded in the theory of superconductivity, as well as in that of nonequilibrium superconductivity, where the energy relaxation rate is a crucial parameter \cite{Kaplan1976}, for example in applications where hot-electron effects are exploited.
In practice, one uses normal-metal or superconducting thin films, in which there is a high degree of impurity scattering dominating the resistivity. 
Thin films with impurity scattering have been extensively studied in the field of weak localization, where the energy relaxation is one of the contributions to the single particle phase-coherence length. 
Quite generally the impurity scattering is considered to be elastic, only changing the direction of momentum, and derivable from the observed resistivity at temperatures above $T_c$. 
This common approach neglects the fact that inelastic electron-phonon scattering is strongly affected  by the presence of impurities or strong disorder. 
Moreover, it overlooks that in many cases of practical interest impurity scattering is predicted to be not elastic but also inelastic, i.e., contributing to the energy relaxation and phase breaking.

In particular, various interference effects modify the temperature dependence of the relaxation rate. 
In impure metals, where the electronic mean free path is less than the wavelength of a thermal phonon, the electron-phonon interaction is found to be suppressed in comparison to the clean case \cite{Pippard1955, Schmid1973, Keck1976, Reizer1986}, and at low temperatures the relaxation rate evolves from the standard $T^3$ dependence in pure metals to $T^4$ in impure metals. 
In this case, the theory assumes that the impurities and defects vibrate in phase with the host atoms of the lattice (the so-called complete drag of impurities). 
More recently, it was pointed out that in the case of even a small difference in the vibrations of the electron scatterers and the host atoms an enhancement of the electron-phonon interaction is expected  \cite{Sergeev2000}. 
Such disorder-enhanced relaxation, with a $T^2$ dependence, was reported for normal metals in measurements of the phase-breaking length in a variety of metallic alloys and summarized by Lin and Bird \cite{LinBird2002}. 
An alternative method to study the inelastic aspects of impurity scattering is by heating the electrons and measuring directly the temperature difference between the electron bath and the phonon bath \cite{Karvonen2005}. 
The applicability of this experimental method to real materials depends strongly on the compatibility with the fabrication technology.

We studied superconducting boron-doped diamond films grown on diamond substrates. An important experimental advantage compared to many previous thin-film studies is that there is no acoustic mismatch between the phonons in the film and in the substrate, and the unified phonon bath should be in equilibrium at the bath temperature. 
Superconductivity in diamond was first found by Ekimov \emph{et al.} \cite{Ekimov2004} in polycrystalline material and by Bustarret \emph{et al.} \cite{Bustarret2004} in single-crystal thin films. 
Boron dopes into a shallow acceptor level close to the top of the valence band that is separated from the conduction band of diamond by $E_g\approx 5.5$ eV. 
At low boron concentrations $n_B\approx 10^{17} - 10^{19}$ cm$^{-3}$ the material is semiconducting. 
If the doping concentration $n_B$ exceeds the critical value ($> 10^{20}$ cm$^{-3}$), the system passes through an insulator-to-metal transition and shows metallic behavior \cite{Klein2007}. 
At a boron concentration of $\ge 5\times10^{20}~$cm$^{-3}$ superconductivity is observed, with an increase in the critical temperature $T_c$ with increasing carrier concentration. 
Crystalline diamond is therefore an attractive model system for the study of the impurity scattering at low temperatures \cite{Sergeev2000}.
The boron dopants provide charge carriers but also play the main role for impurity scattering \cite{Blase2004}. 
Superconductivity has been attributed to the optical phonons arising from the B-C stretching mode\cite{Blase2004}, and these phonons are therefore direcly related to the acceptor atoms. 
Taking into account the absence of a Kapitza resistance for phonons, the CVD-grown boron-doped diamond films form a unique system  to study electron-phonon interaction processes. 
We use amplitude-modulated absorption of (sub-) THz radiation (AMAR), introduced by  Gershenzon \emph{et al.}~\cite{Gershezon1990}, suitable for the study of electron-phonon processes in various thin films, provided that they become superconducting at an experimentally accessible transition temperature.

Our main result is that the relaxation of the resistive superconducting state of boron-doped diamond is controlled by an electron-phonon inelastic scattering rate, which varies as $T^2$. In addition we find that on both sides of $T_c$ the observed relaxation time increases, suggesting a divergent behavior upon approaching $T_c$.   

\section{Determination of the energy-relaxation time in thin superconducting films}

\begin{figure}
\includegraphics{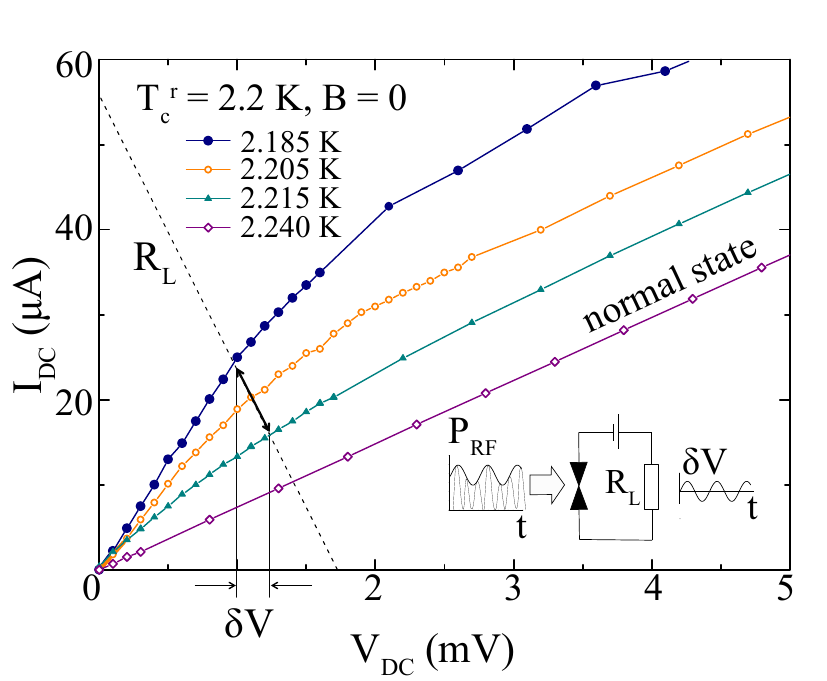}
\caption{\label{fig:fig_d1} Current-voltage curves at different temperatures near the critical temperature $T_c^r$ in zero magnetic field. Under the RF power an operation point ($I$ = 20 $\mu$A and $V$ = 1 mV) shifts along the equivalent load resistance line that produces the voltage signal $\delta V$. The inset shows the equivalent circuit for determining an output signal of the superconducting film upon the absorption of the amplitude-modulated radiation.
}
\end{figure}

The method introduced by Gershenzon \emph{et al.} \cite{Gershezon1990} (AMAR) allows to measure the energy relaxation rate between electrons and phonons. 
A superconducting film is brought into a regime where the film is in a superconducting resistive state (cf. Fig.~\ref{fig:fig_d1}).

\begin{figure}
\includegraphics{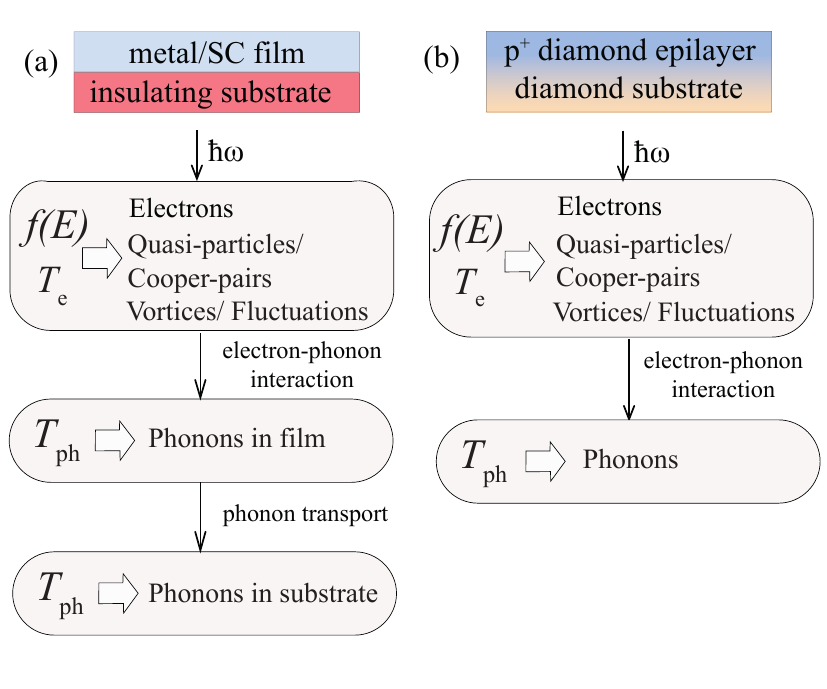}
\caption{\label{fig:fig_d2} Coupling between the thermodynamic subsystems in the case of  (a) a thin metal or superconducting (SC) film on an insulating substrate; (b) a boron-doped diamond film on a diamond substrate, illustrating, in comparison to (a), the absence of a Kapitza resistance. The electron reservoir under illumination can be described by the Fermi-Dirac distribution function $f(E)$ with an effective electron temperature $T_e$ exceeding the phonon temperature $T_{ph}$. In practice, the metal film is in a resistive superconducting state which is due to vortices, phase and/or amplitude fluctuations. It is assumed that the electron temperature $T_e$, increased by DC power and by RF power, controls the changes of the resistivity of the superconducting state.}
\end{figure}

 A small dc current $I_{DC}$ is applied and the voltage $V_{DC}$ is monitored, while an amplitude-modulated signal from a submillimeter source is directed at the film. The modulation frequency for the amplitude is $\omega_m$. 
The absorbed radiation power causes an increase of the electron temperature $T_e$ (cf. Fig.~\ref{fig:fig_d2}), which leads to an increase of the film resistance $\delta R$ followed by a voltage signal proportional to the bias current $\delta V = I\delta R$. 
It is found that this method provides a frequency-dependent rolloff (cf. Fig.~\ref{fig:fig_d3}), which is taken as the measure of the energy-relaxation rate.
 Application of a perpendicular magnetic field brings the film into a resistive state at various temperatures, which makes it possible to measure the energy-relaxation rate  as a function of temperature, although over a limited range of temperatures. 
 The technique has been applied to various materials, usually providing different  temperature dependences of the relaxation rate, for example $T^2$ in Nb film \cite{Gershezon1990}, $T^{1.6}$ in NbN \cite{Gousev1994} and $T^3$ in TiN \cite{Kardakova2013} and in NbC \cite{Ilin1998}. 
 
Inelastic relaxation times are important in nonequilibrium superconductivity, in hot-electron effects in normal metals, and in magnetoresistance due to quantum-interference processes. 
In the latter, the normal-metal phase-breaking time $\tau_\phi$ is limiting the phase coherence of elastically scattered electron waves. 
It is usually assumed to be limited by inelastic electron-phonon scattering and at lower temperatures by electron-electron scattering.
 Gershenzon \emph{et al.}~\cite{Gershezon1990} have made a comparison of results based on weak-localization experiments with those obtained with the AMAR method. 
 The outcome, for strongly disordered niobium films with the elastic mean free path in the 1-nm range and a $T_c$ of 3.2 to 8.5 K, is that the results are comparable in the temperature range from 10 to 20 K. 
 At lower temperatures the weak-localization results differ quite strongly, both in the temperature dependence and in the absolute value, which is attributed to electron-electron scattering as the limiting process for weak localization.
It supports the assumption that an elevated electron temperature can be assigned to the electron system   in comparison to the phonon-temperature.
 Although the complexities of electron-electron scattering, in particular in relation to its material dependence, have meanwhile become much more detailed ~\cite{Pothier1997, Gougam2000, Pierre2003, Anthore2003,Huard2005}, we assume that the basic premise of the existence of an electron temperature is justified.  

A boron-doped diamond film is considered a thermodynamic system  composed of two interacting subsystems: electrons and phonons which are coupled via electron-phonon interaction \cite{Wellstood1994}. 
If the film is exposed to amplitude-modulated radiation, the temperature of the electron subsystem will change accordingly.
 The amplitude of the temperature response will depend on the modulation frequency and the time constant of the electron subsystem. 
 This time constant will be a function of the electronic specific heat and the heat conductance between the electrons and the phonons in the film.
  Due to the absence of a film-substrate interface for phonons between the doped layer and the substrate we assume that the escape time of nonequilibrium phonons into the substrate is very short, considerably less than the resistance-relaxation times we find. 

The resistance relaxation time is determined from 3-dB rolloff of the frequency dependence of the amplitude of the output voltage $\delta V(\omega_m)$. Since the diamond film has no Kapitza resistance, the phonons can be treated as a heat bath in equilibrium with the cryogenic environment. Then the dynamics of the film can be described by a single heat-balance equation:

\begin{equation}
C_e \frac{d T_e}{dt} = -G(T_e - T_b) + P_{DC} + P_{RF},
\label{eq1_heat}
\end{equation}
with $C_e$ the heat capacity of the electrons, $T_e$ the electron temperature,  $T_b$ the phonon-bath temperature, $G$ the heat conductance from the electrons to the phonon-bath, $P_{DC} = I^2R$ the Joule power dissipated in the film, and $P_{RF}$ the absorbed radiation power. Equation (\ref{eq1_heat}) is valid under two assumptions. The first one is that the Joule heating and the RF drive are sufficiently weak so that the departure of $T_e$ from $T_b$ is small in the sense $|T_e - T_b|\ll T_b$. This regime is achieved when the distance $L$ between the contact pads is larger than the thermal diffusion length $L_{diff} = \sqrt{D\tau_{e-ph}}$ (with $D$ the electronic diffusion coefficient and $\tau_{e-ph}$ the electron-phonon interaction time). As will be shown below, the condition $L_{diff} \ll L$ is satisfied in our case. In the experiments the radiation power was modulated, so that $P_{RF}(t) = P_0 + P_1 exp (i\omega_m t)$. This allows us to use the results of the lumped-circuit model for hot-electron bolometers given by Karasik and Elantiev \cite{Karasik1996} for the power of the response signal generated by the  film: 

\begin{equation}
P_{out}(\omega_m) = \frac{P_0}{1+(\omega_m\tau_B)^2},
\label{eq2_power}
\end{equation}
with $P_0$ the power for low modulation frequency and 

\begin{equation}
\tau_B = \frac{\tau_{e-ph}}{1+\alpha}, \text{with}~ \alpha = \frac{I^2}{G}\frac{\partial R}{\partial T_e}\frac{R - R_L}{R_T + R_L}.
\label{eq3}
\end{equation}

In Eq.~(\ref{eq3}), $R = V/I$, is the Ohmic resistance at the operating point, and $R_L$ is the equivalent load resistance determined by the read-out electronics and the bias circuit, $R_T = R + I(\partial R/\partial I)$. By plotting $P_{out}(\omega_m)$ we determine the time constant $\tau_B$. Choosing the operating point so that the current is as small as possible, while the response power is still measurable, allows minimizing the parameter $\alpha$. Physically, in this regime the Joule heating is minimal and is assumed not to slow down the energy relaxation process of the electron system. Keeping the current low also reduces nonthermal effects such as vortex-creation or enhanced phase-slip rates, which are not included in Eq.(\ref{eq1_heat}). 

An equivalent electrical circuit for analzing the response of the superconducting film to amplitude-modulated radiation is presented in the inset of Fig.~\ref{fig:fig_d1}. As a radiation source, we use a backward-wave oscillator (BWO) with a carrier frequency of 350 GHz. The BWO power is amplitude-modulated at frequencies of 10 to 2000 kHz. The response voltage from the film,  $\delta V(\omega_m)$, and the frequency $\omega_m$ are measured with a spectrum analyzer.  To determine the temperature dependence of the resistance-relaxation time we varied the bath temperature and applied a magnetic field to get into a usable resistive superconducting state by the creation of vortices. 

A crucial ingredient of this AMAR method is the exploitation of the resistive superconducting state. The observed relaxation is in essence the restoration of the resistive superconducting state after exposure to radiation with a frequency higher than the energy gap of the superconductor. For the results presented here we use, in practice, two types of resistive superconducting states. 

Case A: the resistive transition of a superconducting film in zero magnetic field. Above the mean field critical temperature $T_c$, the resistive transition is determined by amplitude fluctuations of the order parameter, and known as Aslamazov-Larkin \cite{Aslamazov1968, *AL1968} and Maki-Thompson contributions \cite{Maki1968, Thompson1970}.  Below $T_c$ it is, for one-dimensional superconductors, determined by thermally activated phase slip events. For a 2-dimensional film the situation is more complex. There exists a well-defined regime for films with a high resistance per square, where the emergence of resistivity is controlled by the theory of the Berezinskii-Kosterlitz-Thouless (BKT) model, which focuses on the macroscopic phase fluctuations. In this theory a superconducting film upon approaching $T_c$ will first pass another critical temperature $T_{BKT}$,  where vortex-antivortex pairs unbind, providing free vortices. These free vortices will move under the influence of a transport current and will therefore provide a voltage across the superconductor and makes it appear resistive. The temperature dependence of the resistivity, the exponential rise at the resistive transition, is due to the increasing presence of free vortices. The relevant temperature is the electron temperature, since it controls the superconducting properties including the density of free vortices. As shown by Kamlapure \emph{et al.} \cite{Kamlapure2010} a detailed analysis of the resistive superconducting properties in terms of the BKT theory can be made, for example for NbN, provided finite-size effects of the films are properly taken into account. For the diamond films studied here, with a not too high sheet resistance, the expected $T_{BKT}$ is close to $T_c$, which rules out such an analysis based only on the phase.  In the regime where $T_{BKT}$ is close to $T_c$, the emergence of resistance is then not exclusively controlled by the vortex density occurring in the BKT theory, but includes both phase and amplitude fluctuations of the order parameter. Therefore, the observed resistance is most likely due to the interplay of the time-dependent phase differences and non-equilibrium conversion currents, as was studied experimentally by Carlson and Goldman \cite{Carlson1976, Carlson1975, Carlson1973}. Because of this complexity a quantitative description of the emergence of resistance in a two-dimensional superconducting film cannot be based on a sharply delimited conceptual framework~\cite{Konig2015}.  It is known that for uniform systems in the limit of $\Delta \ll k_B T$ the relaxation of nonequilibrium state induced by radiation should be called longitudinal nonequilibrium \cite{Tinkham}. Its relaxation is controlled by the electron temperature $T_e$, by the mean-field critical temperature $T_c$, and the inelastic relaxation rate $\tau_E$. In many cases $\tau_E$ is the elecron-phonon time, which itself is temperature dependent. In addition there is the temperature dependence related to the restoration of the superconducting state, which is dependent on $(1 - T_e/{T_c})^{-1/2}$. 

In practice a dc bias is also used to move the resistive transition to lower temperatures, which allows a range of temperatures close to $T_c$ to be accessed. The shift of the resistive transition is due to the fact that a transport current also contributes to the creation of  extra free vortices in the BKT theory \cite{Halperin1979}.  

Case B: the resistive state of the superconductor is reached by applying a perpendicular magnetic field, which creates vortices, with their flux in the direction of the applied magnetic field above the field $B_{c1}$. With a current applied these vortices move under the Lorentz force, provided the force exceeds the pinning force. This flux-flow regime including the breakdown of collective flux-pinning was studied recently for NbN films by  Lin et al. \cite{Lin2013}. Case B enables carrying out measurements over a larger range of temperatures. By choosing a bath temperature $T_b$ and adjusting the magnetic field close to $B_{c2}$ at that temperature, we access the temperature-dependent resistive superconducting state. By appying a low bias-current, we choose a resistive state where a measurable voltage response exists. Also in this case the resistive superconducting state is controlled by the electron temperature. In contrast to case A the superconducting state is in principle in the regime $\Delta\gg k_B T_c$.  In this way it is possible to measure a voltage from the resistive superconductor, caused by the modulation of the electron temperature as a function of applied modulation frequency at different bath temperatures. The main assumption is that the changes of the resistive superconducting state, caused by flux flow,  with absorbed power are due to a rise in electron temperature and do not contain any corrections due to the fact that a magnetic-field-induced resisitive \emph{superconducting} state is used. In other words, the fact that the resisitivity is due to flux-flow processes and by the density of vortices is not affecting the observations in a significant way. The only significant parameter is the effective electron temperature for a given B-field and the current, in comparison with the phonon-bath temperature. In addition, it is experimentally verified that the observed response does not depend on the level of the microwave power (linear regime in power).

\begin{table*}
\caption{\label{tab:table1}Parameters of the films}
\begin{ruledtabular}
\begin{tabular}{cccccccc}
 Sample & $d$ & $R_\square$ & $\rho$ & $T_c^r$ & $D$ & $\alpha$ & $p$\\
 & (nm) & (Ohm)& ($\mu$Ohm$\times$cm) & (K) & (cm$^2$s$^{-1}$) & ($\mu$s$\times$K$^p$) & \\
 \hline
 N1 & 300 & 50 & 1500 & 2.245 & 1.38$\pm$0.04 & 1.91 & 1.88$\pm$0.05 \\
 N2 & 70 & 220 & 1540 & 2.195 & 1.30$\pm$0.02 & 1.92 & 2.06$\pm$0.05 \\
 \end{tabular}
\end{ruledtabular}
\end{table*}

With the assumptions stated above the temporal response of the resistive superconducting state, which we observe in the experiment, serves in all cases as a measure of the temporal response of the electron temperature. For the regime of the time constants that we find, this seems like a justified assumption. However, since we are observing the resistivity of the superconducting state the restoration of the superconducting state adds in principle an additional temperature dependence around $T_c$.  

\section{Samples}

Two $p^+$ epilayers of diamond were grown in a home-made vertical silica tube reactor \cite{Achatz2010} by microwave plasma-enhanced chemical vapor deposition (MPCVD) on $0.3\times3\times3$ mm$^3$ (001)-oriented type Ib diamond substrates, on top of a 500-nm-thick nonintentionally doped buffer layer. The growth was carried out at $880~^{\circ}$C in a gas mixture of H$_2$, CH$_4$, and B$_2$H$_6$. The total pressure was 33 torr, i.e. 44 hPa, and the total gas flow was 100 sccm. The molar methane-to-hydrogen ratio was $3.5$~\%, and the boron-to-carbon molar ratio in the introduced gas mixture was 0.25~\% for sample N1 and 0.33~\% for sample N2. The other difference between the two samples was the duration of the growth, leading to thicknesses $d$ of 300 and 70 nm for samples N1 and N2, respectively.
Four parallel silver-paste contacts drawn across the whole sample were used to measure the sheet resistance $R_{\square}$ of the film at 300 K, and the thickness was deduced from spectroscopic ellipsometry measurements \cite{Bousquet2014} performed \emph{in situ} \cite{Fiori2014}. This allowed the determination of the resistivity $\rho = R_{\square} d$, which was very similar for both samples. The critical temperature $T_c^r$ was determined as the temperature of the midpoint of the resistive transition where the sample's resistance is 50~\% of $R_N$ (the $R_N$ is the normal state resistanse above the transition). We also measured the electron diffusion constant $D$ from the temperature dependence of the second critical magnetic field $H_{c_2}$ as

\begin{equation}
D = \left.-1.28 \frac{k_B c}{ e}\left(\frac{dH_{c_2}}{dT}\right)^{-1} \right|_{T=T_c}
\end{equation}

The results are summarized in Table~\ref{tab:table1}. 

\section{Three regimes}

\begin{figure}
\includegraphics{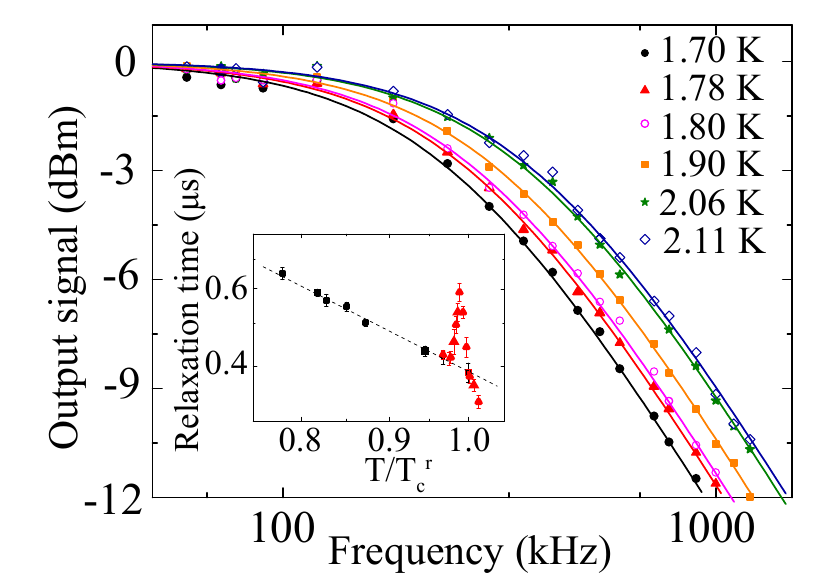}
\caption{\label{fig:fig_d3} The frequency dependence of the sample response at different bath temperatures $T_b$. The experimental data are measured at a temperature in the middle of the superconducting transition (where $\partial R/\partial T = $ max) at the same bias current. The temperature of the resistive transition shifts when a magnetic field normal to the film plane is applied. The data of each curve were normalized to 0 dB for convenience. The solid lines are a least-square fit with Eq.~\ref{eq2_power}. The fit standard error of the rolloff frequency does not exceed 10 \%. The inset shows the energy relaxation time vs the normalized temperature ($T/T_c^r$), where the critical temperature $T_c^r$ is    the temperature of the middle point of the resistive transition. The experimental results correspond to both types of measurements: case A (red triangles) and case B (black squares). 
}
\end{figure}

A typical experimental result is shown in Fig.~\ref{fig:fig_d3} for one of the samples (sample N2). This set of data is taken following the case-B method described above. The bath temperature is set between 1.7 K and 2.11 K. A perpendicular magnetic field is applied until a resistive state is reached, providing the resistive state shown in Fig.~\ref{fig:fig_d1}. The output voltage $\delta V(\omega_m)$, as a function of modulation frequency is shown in Fig.~\ref{fig:fig_d3}.  With increasing frequency we find for each bath temperature a clear rolloff. We apply a least-square fit to the measured data using Eq.~(\ref{eq2_power}), which leads to a characteristic relaxation time, shown in the inset. Similar curves are obtained for measurements using the case-A method. All the results are put together in the inset of Fig.~\ref{fig:fig_d3} as a function of the normalized critical temperature $T_c^r$, determined from the midpoint of the transition in the absence of a magnetic field which we call $T_c$ of the film, as listed in the Table~\ref{tab:table1}.

We identify three different regimes in the response: 

\begin{itemize}
\item{Regime I: At temperatures $0.75~T_c < T < 0.95~T_c$, the data for both samples are shown in Fig.~\ref{fig:fig_d4}. We observe, for both samples, a very similar trend  with $\tau = \alpha T^{-p}$, with $\alpha$ and $p$ used as fitting parameters.  The values of $\alpha$ and $p$ obtained from the least-square fit are listed in Table ~\ref{tab:table1}.In both cases the value of $p$ is very close to 2. The values of $\tau (T)$ run from  400 to 700 ns over the temperature range 1.7 K to 2.2 K.} 

\item{Regime II: At temperatures $0.95~ T_c < T < 0.99~ T_c$, the relaxation time increases in a divergent manner upon approaching $T_c$ (inset of Fig. ~\ref{fig:fig_d3}). This is reminiscent of data reported before by Gershenzon \emph{et al.} \cite{Gershezon1984} for dirty niobium and interpreted as the observation of the relaxation of the superconducting order parameter, the so-called longitudinal relaxation time known from non-equilibrium superconductivity \cite{Tinkham}.} 

\item{ Regime III: At even higher temperatures of $0.99~T_c < T < 1.02~ T_c$, the relaxation time decreases with temperature. In this regime the resistive superconducting state is close to the normal-state resistance. It should be considered as within the range where the superconducting state emerges out of the normal state due to time-dependent Ginzburg-Landau fluctuations, i.e., within the resistive transition. } 
\end{itemize}

These three regimes represent in our view three different physical processes. We consider the fact that regime II and III have been measured according to the case A method and regime I with the case B method an important distinction of which the significance is to be addressed further. In case A we are in a regime where many processes are entangled and where one can safely state that $\Delta \ll k_B T\sim k_B T_c$. In case B there is a well developed energy gap $\Delta$ outside the regime where the vortex-cores are located, but the magnetic field is close to $B_{c2}$. Therefore the resistive superconducting state is controlled by a complex inhomogeneous nonequilibrium process.  The restoration of the resistive state occurs in a spatially distributed way with, on a microscopic level, scattering and recombination processes known from nonequilibrium superconductivity, as well as diffusion processes. 

\section{Regime I}

\begin{table*}
\caption{\label{tab:table2}Calculated parameters of the films for evaluation of $\tau_{e-ph} (T)$}
\begin{ruledtabular}
\begin{tabular}{ccccccc}
 Sample & $n_B$ & $\epsilon_F$ &  $k_F$ & $l$ &  $N_0$ &  $b$\\
 & (cm$^{-3}$) & (eV)& (cm$^{-1}$) & (nm) & (eV$^{-1}\mu$m$^{-3}$) & \\
 \hline
N1 & 3$\times$10$^{21}$ & 1.5 & 4.5$\times$10$^7$ & 0.41   & 3.1$\times$10$^9$ & 0.066 \\
N2 & 3$\times$10$^{21}$ & 1.5 & 4.5$\times$10$^7$ & 0.39  & 3.1$\times$10$^9$ & 0.073 \\
 \end{tabular}
\end{ruledtabular}
\end{table*}

\begin{figure}
\includegraphics{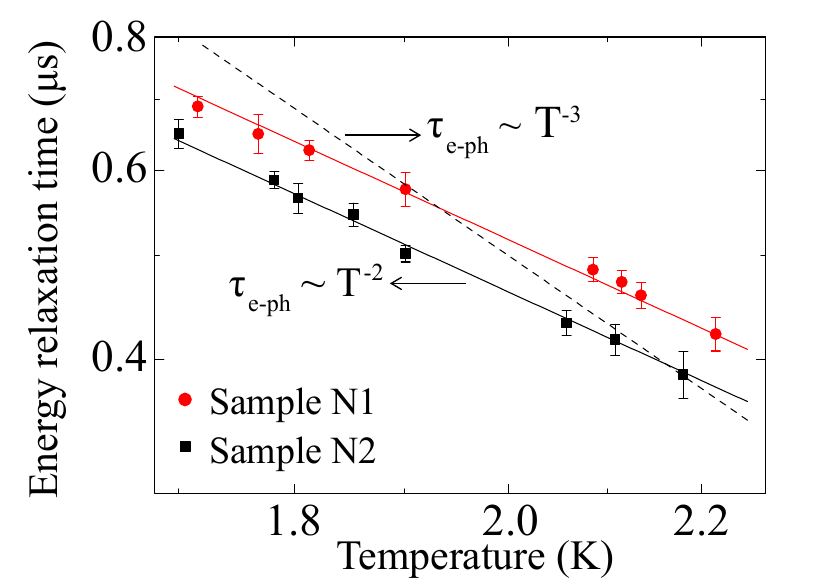}
\caption{\label{fig:fig_d4} The temperature dependence of the energy-relaxation time at low temperatures far from $T_c$ ($0.75 T_c^r < T < 0.95 T_c^r$). The full lines represent the theoretical estimate for $\tau_{e-ph}(T)$ according to Eq.~(\ref{eq:T^2}). For comparison the dashed line represents a $T^{-3}$ dependence.
}
\end{figure}

The observed resistance relaxation time indicates how fast the resistance changes with a modulation of the input power. As argued above we interpret this time as the energy-relaxation time between the electron and the phonon system. In previous experiments, such as for TiN \cite{Kardakova2013}, we find that it obeys a power law with the exponent $p = 3$. Here we find clearly the exponent $p = 2$ (Fig.~\ref{fig:fig_d4}). 

A straightforward explanation for $p = 2$ might be the dimensionality. A phonon system is two-dimensional when $\lambda_T \gg d$, with $\lambda_T$ the wavelength of the thermal phonons and $d$ the film thickness. This is definitely not our case because the wavelength of the thermal phonon $\lambda_T = (\hbar u_l)/(k_B T)$ is $\approx$ 60 nm, which is less than the thicknesses of both our samples. Besides, since our superconducting layer is grown on diamond the perfect acoustic match between the boron-doped diamond epilayer and the diamond substrate makes it unreasonable to think in terms of two-dimensional phonons. 

In the case of a three-dimensional phonon system substantial modifications in the electron-phonon interaction due to electron-impurity scattering have been developed. They depend on the polarization mode, transverse or longitudinal phonons, and on the effect of disorder. Furthermore, the samples we study are in the dirty limit in the sense $l\ll \lambda_T$, which is already achieved at $T\approx 2.2$ K with $l/\lambda_T = 0.2$.

In disordered metals the electron-phonon interaction is non-local with a characteristic size of the interaction region about equal to $\lambda_T$. In the diffusive limit, when $l \ll \lambda_T$, the theory predicts, in the presence of strong elastic scattering,  a weakened electron phonon interaction \cite{Schmid1973, Keck1976, Reizer1986}:

\begin{equation}
\tau_{e-ph} = \frac{1}{9.1} \frac{10}{3\pi^4 \beta_t}\frac{(p_Fu_t)^3}{p_F l }\frac{1}{(k_B T)^4},
\end{equation}
with $\beta_t =(2\epsilon_F/3)^2 (N_0/(2\rho_m u_t^2 ))$ the coupling constant, $p_F$ and $\epsilon_F$ the Fermi momentum and Fermi-energy, $N_0$ the density of states at the Fermi energy and $\rho_m$ the mass density (these parameters are listed in Table~\ref{tab:table2}). The coefficient 9.1 results from averaging over all electron states contributing to $\tau_{e-ph}$ \cite{Ilin1998}. This theoretical model assumes that all the impurity scattererers for electrons, vibrate in phase with the host atoms. Experimentally, the $T^{-4}$ behavior of $\tau_{e-ph}$ has been predominately observed in elemental thin films, such as Cu \cite{Karvonen2005}, Au \cite{Karvonen2005}, Hf \cite{Gershenson2001} and Ti \cite{Gershenson2001}, mostly at very low temperatures, below hundreds of mK. Nevertheless, the $T^{-4}$ dependence was also found in disordered amorphous InO films \cite{Sacepe2009} and heavily doped silicon \cite{Kivinen2003} at low temperatures. Obviously, the results found in our diamond films are not in agreement with this model for electron-phonon interaction in the presence of impurity scattering.  

A $T^{-2}$ dependence, found from weak localization experiments, has been reported for alloys, polycrystalline films, and metallic glasses such as CuZrAl \cite{Li2006}, TiAl \cite{DiTusa1992}, TiAlSn \cite{Lin1995}, AuPd \cite{Hsu1999}, VAl alloys \cite{Zhong1998}, CuCr \cite{Meikap2004}, ZrSn, Au-doped In$_2$O$_{3-x}$ films \cite{Ceder2005} and in Mn-doped Al films \cite{Underwood2011}. Such disorder-enhanced relaxation, with a $T^{-2}$ dependence, is predicted by a recent model of Sergeev and Mitin (SM) of scattering of electrons by static impurities such as heavy impurity atoms, the defects, and grain boundaries \cite{Sergeev2000}: 

\begin{equation}
\tau_{e-ph} = \frac{1}{1.6} \frac{1}{b} \frac{(p_F l)(p_F u_t)}{3\pi^2 \beta_t }\frac{1}{(k_B T)^2},
\label{eq:T^2}
\end{equation}
where the coefficient $b$ ($b_{max} = 0.25$)\cite{Karasik2012} describes the difference in the vibration of the scatterers and the host atoms. We apply this theoretical prediction to our data as follows.  
The carrier density is given by $n_B = 3\times 10^{21}$ cm$^{-3}$ from the experimental dependence of the critical temperature as a function of the boron concentration \cite{Klein2007}. The Fermi wave vector and the elastic-scattering length are determined from $k_F = \sqrt[3]{3\pi^2 n}$ and $l = ((3\pi^2)^{1/3} \hbar) /(3 e^2 \rho \pi^2 n^{2/3})$ within the Drude-Sommerfeld model \cite{Ashcroft}. The effective carrier mass follows from  $ m^* = (p_F l)/(3D)\approx 0.5 m_e $, where $m_e$ is the electron mass. The density of states at the Fermi level is estimated from the experimental values of the resistivity $\rho$ and the electron diffusion constant $D$ through the expression $ N_0 = 1/(e^2 \rho D)$. The mass density $\rho_m$ was taken for diamond with the value $\rho_m = 3.5$ g/cm$^3$. The sound velocities for the longitudinal mode $u_l = 16\times 10^5$ cm/s and for the transverse mode $u_t = 9.7 \times 10^5$ cm/s are estimated from the phonon dispersion relations using Giustino \emph{et al.}~\cite{Guistino2007}.  The calculated dependencies are shown in Fig.~\ref{fig:fig_d4}, using only the parameter $b$  as a fitting parameter (see Table~\ref{tab:table2}).

Since the elastic mean free path $l$ of electrons is comparable with average distance between the boron atoms, we assume that the carriers are scattered predominantly at sites of boron atoms. However, the mass difference between the boron and carbon is only about 10\%, which itself is not a sufficient condition for the applicability of the SM model. Therefore in the case of boron-doped diamond one should consider as  scatterers also clusters of boron atoms (dimers, trimers and etc.)\cite{Bourgeois2006}, but further studies are needed to identify the exact nature of the scatterers.  

For completeness we point out that a $T^{-2}$ dependence of the relaxation time is also predicted for semiconductors for the case of low screening \cite{Sergeev2005}. The e-ph interaction (through the deformation potentials) in semiconductors is of different nature than those for metals\cite{Gantmakher1987}. This interaction in semiconductors has a different dependence on disorder and on the electronic concentration. However,  because of the relatively high boron concentration (of order $10^{21}$ cm$^{-3}$), our diamond samples are in the strong screening limit (with the screening length $\kappa^{-1}\approx 1.5$ \AA, where $\kappa^2 = 4\pi e^2 N_0$), and hence the theory for e-ph interaction in a semiconductor is not applicable to this case. 

It should also be stressed that the time $\tau_{e-ph}$, which is measured with the AMAR technique is by definition the time of relaxation of the electron temperature due to electron-phonon interaction. It differs from another characteristic electron-phonon time - the time of the relaxation of the distribution function - by a numerical factor and is several times shorter. The reason for this difference is the following. The rate of relaxation for a quasiparticle depends on its energy and increases with it. Thus the total energy of the thermal distribution of quasiparticles (and correspondingly the temperature), which is determined mainly by quasiparticles with the highest energies, relaxes faster than the number of quasiparticles. The value of the numerical factor depends on the particular form of the electron-phonon collision integral and has not been calculated for the most general case~\cite{[{For the conventional case of electron-phonon interaction in clean limit, it is 4.5, see in }] Allen1987}. Because the time of the relaxation of the distribution function, or of the quasiparticle number, is also often referred to as the electron-phonon time, one should bear in mind the numerical difference between it and the time of the relaxation of the electron temperature. But, the temperature dependences for both are the same. The formulas (see above) we will use to fit the experimental data give the time of the relaxation of the quasiparticle number, not the time of the energy relaxation, but because of an uncertainty in the numerical coefficient in these formulas and of an unknown factor between the two times, we will neglect this difference.

\section{Regime II} 

At the temperatures in the range $0.95 T_c < T < 0.99 T_c$, the relaxation time is found to increase sharply (Fig. \ref{fig:fig_d3}). In this temperature range the photons of the THz source scatter quasiparticles to energies well above the superconducting energy gap at the given temperature. The resistive superconducting state is expected to relax back to the equilibrium state on a time scale of the order of the so-called longitudinal relaxation time, Eq.~(\ref{eq:tau_L}).  

\begin{figure}
\includegraphics{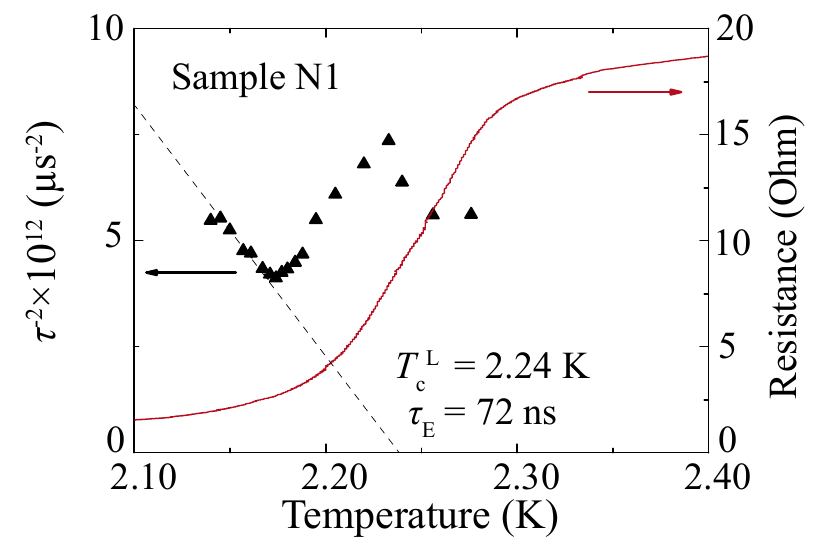}
\includegraphics{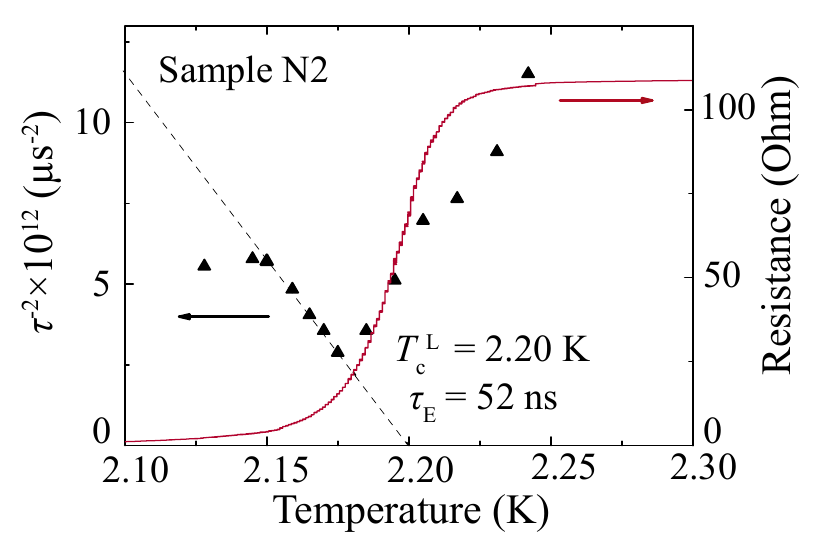}
\caption{\label{fig:fig_d5} The inverse square of the relaxation time ($\tau^{-2}$)  and the film resistance as function of the temperature. The dashed line corresponds to the longitudinal relaxation time, which diverges as $(T_c^L/(T_c^L - T))^{1/2}$, where $T_c^L$ is determined as the temperature at which the superconducting gap is completely suppressed and $\tau^{-2} = 0$. The values of $T_c^L$, indicated in the legend, are close to the temperature $T_c^r$ determined from the resistive transition. 
}
\end{figure}

Since the longitudinal relaxation time is inversely proportional to the energy gap, we plot the data,  Fig.~\ref{fig:fig_d5},  as the inverse square of the relaxation time vs temperature. We obtain straight lines suggesting that we are indeed observing  the longitudinal relaxation time, which diverges as $(T_c/(T_c-T))^{1/2}$, although the extrapolated value goes to a $T_c$, which we called $T_c^L$ which is slightly different from $T_c^r$. This longitudinal relaxation time is given by: 

\begin{equation}
\label{eq:tau_L}
\tau_L \approx 3.7 \tau_E k_B T_c / \Delta
\end{equation}
with $\Delta (T \approx T_c) \approx 3.1 k_B T_c (1-T/T_c )^{1/2}$, where $T_c$ is the critical temperature, i.e., the temperature at which the gap is completely suppressed, and $\tau_E$ is the energy-relaxation or inelastic-scattering time for an electron at the Fermi surface. In this case the critical temperature $T_c$ is determined as the temperature $T^L_c$ at which the value of the order parameter approaches zero. The values of $T^L_c$ for both samples are almost identical to the values of $T^r_c$ determined from the resistive transition.

  The time $\tau_E$ is the characteristic time for the nonequilibrium distribution function to relax to the Fermi function. In the standard analysis of, for example Kaplan \emph{et al.}~\cite{Kaplan1976}, this $\tau_E$ is related to electron-phonon interactions as measured in energy-dependence of the superconducting energy gap in a tunneling experiment. Hence, the inelastic scattering rate is coupled to the electron-phonon interaction responsible for superconductivity. However, in general,  two processes may be responsible for inelastic scattering: electron-electron interaction and electron-phonon interaction. The faster of the two will dominate. The estimated values of $\tau_E$ at $T_c^L$ are $\tau_E \approx 52$ ns for the 70-nm sample (with $T_c^L = 2.2$ K) and $\tau_E  \approx 72$ ns for the 300-nm sample (with $T_c^L = 2.24$ K). 
While comparing the inferred values of $\tau_E$ to the characteristic time of electron-phonon interaction, one should remember that $\tau_{e-ph}$ measured in regime I is the time of the energy relaxation and should be several times shorter than the time of the relaxation of the distribution function in the same process. Thus $\tau_E$ is more than an order of magnitude less than the time of relaxation of the distribution function due to electron-phonon interaction.

In a pure metal  the relaxation time due to electron-electron interaction follows, for states near the Fermi surface, an inverse quadratic temperature dependence:

\begin{equation}
\tau_{ee}\propto \frac{\hbar \epsilon_F}{(k_B T)^2}
\end{equation}

The values of $\tau_{ee}$ in the clean case are of the order of 15 ns at $T_c=2.25$ K. In dirty metals with a short electronic mean free path the electron-electron interaction is enhanced compared to the clean case. The  actual relation depends on the dimensionality, which in turn depends on the ratio of the film thickness to the characteristic length $L_T = \sqrt{\hbar D/ (k_B T)}$ called the thermal diffusion length. It defines the length scale over which electrons loose coherence as a result of the thermal smearing of their energy \cite{Altshuler1980}. Since for our samples $L_T \approx 25$ nm, both samples are in the three-dimensional regime for electron-electron interaction. 

In 3-D dirty metals, for an electron at the Fermi surface, the electron-electron scattering rate \cite{Schmid1974, Rammer} is given by 

\begin{equation}
\frac{1}{\tau_{ee}}=c\frac{1}{k_F l}\frac{(k_B T)^{3/2}}{\epsilon_F \sqrt{\hbar\tau}}
\end{equation}
where $c =(3\sqrt{3\pi})/16 \zeta(3/2)(\sqrt{8}-1) \cong 2.75$, and $\tau = l^2/ 3D$ is the elastic scattering time. For $T_c = 2.25$ K we obtain $\tau_{ee}\approx 0.2$ ns. This value is considerably less than the experimentally determined values of $\tau_E$. Therefore we believe that the electron-electron interaction does not play a role in the interpretation of the data in this regime. 

We conclude this section by emphasizing that the relaxation time of 500 ns found in regime I is different from the relaxation time of 50 to 70 ns found in regime II. We note however, that the unprocessed relaxation times measured in case A and case B are quantitatively at the same level. Therefore we assume that our identification of the data obtained in case B as representing the bare $\tau_E$ and the ones of case A the longitudinal relaxation rate is too simplified can not be used too strongly for the absolute value. In reality the restoration of the resistive state in case B involves an inhomogeneous state with unipolar vortices and elsewhere a well-developed energy gap, although close to $B_{c2}$. In case A we deal with a system very close to $T_c$ also inhomogeneous and in the limit where $\Delta \ll k_B T$ and where multiflux-quantum domains may exist with opposite polarity. The restoration of the resistive superconducting state in the time domain involves a complex process, which may influence the absolute values. We believe however, that we can safely attribute significance to the observed $T^{-2}$  and $(T_c/(T_c-T))^{1/2}$-dependences in comparison with other superconducting materials.   

\section{Regime III}

Finally, at higher temperature in the range  ($0.99T_c < T < 1.02 T_c$) the resistance-relaxation time falls with temperature. The decrease of $\tau$ corresponds to the temperature region of the conventional resistive transition, where thermally activated processes generate vortices/phase slips, as well as amplitude fluctuations, which gradually merge towards the regime of superconducting fluctuations out of the normal state.  Above the superconducting mean field $T_c$ the fluctuations can be described by the Aslamazov-Larkin theory using the time-dependent Ginzburg-Landau (TDGL )equations\cite{Tinkham}. 

\begin{figure}
\includegraphics{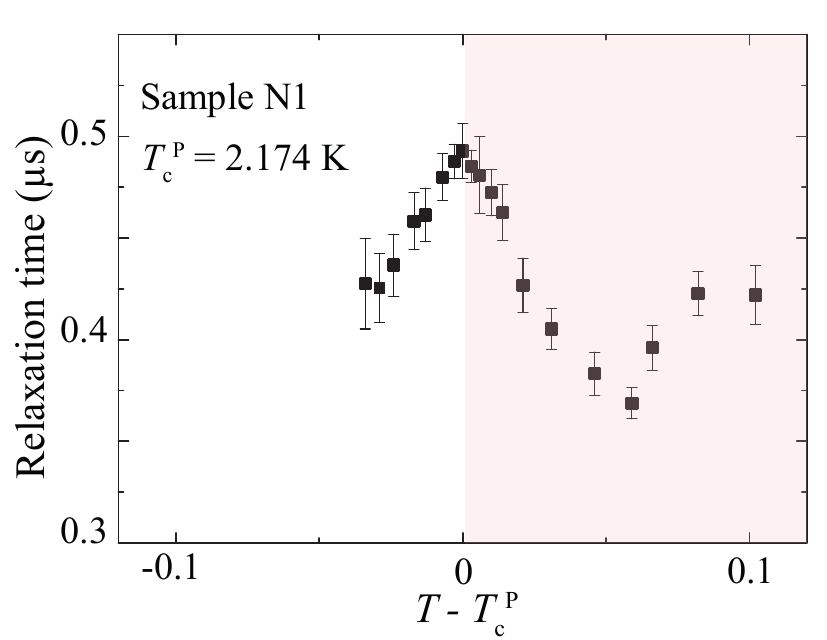}
\includegraphics{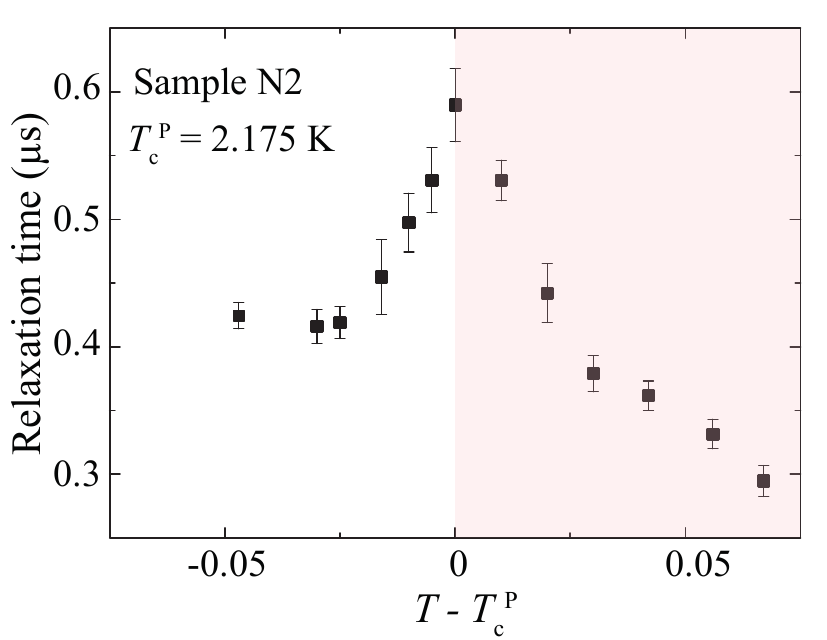}
\caption{\label{fig:fig_d6} The relaxation time \emph{vs.}  $T - T_c^p$, where the temperature $T_c^p$ correspond to a temperature at the peak separating regime II and regime III. On both sides of the temperature $T_c^p$ the observed relaxation time increases, suggesting a divergent behavior upon approaching $T_c^p$. For Sample N1, since the last 3 points were measured in the limit where the film is normal, we left them out of the discussion.
}
\end{figure} 

According to this time-dependent Ginzburg-Landau theory the characteristic time is controlled by

\begin{equation}
\tau_{0}=\frac{\pi\hbar}{8k_B (T-T_c)},
\end{equation}
which is a measure of how quickly a temporary existence of superconducting order gets restored to the normal state. As shown in Fig. ~\ref{fig:fig_d6}, for descending temperatures upon approaching $T_c$ the lifetime of superconducting order gets extended in order to become 'infinitely' long. The temperature located at the peak between regime II and regime III, is denoted as $T_c^p$. The values of $T_c^p$ differ from the values of $T_c^r$ and $T_c^L$ within 1\% - 3\%. This difference in temperatures may be due to the $T_c^r$ having been measured with minor disturbance, whereas $T_c^L$ and $T_c^p$ both represent bias-conditions with a sizable measurement current. The temperature $T_c^p$ can be considered as the superconducting mean field $T_c$ above which the fluctuations of the order parameter dominate. The observed temperature dependence $\tau(T)$ is in agreement with such a scenario (Fig.~\ref{fig:fig_d6}), but the data are too limited to conclude that this is indeed what we observe. Regime III has not been reported before with this experimental method. In comparison with previous measurements, where the spontaneous thermal fluctuations of the order parameter were measured with a tunnel junction ~\cite{Anderson1972, Skocpol_1975}, the advantage of our method is a direct restoration of the electron system after a disturbance and a possibility to maintain phonons in equilibrium. However, more detailed measurements are needed to resolve the situation more accurately. Nevertheless, it is to be expected that on both sides of the mean-field critical temperature $T_c$ we will have a divergent slowing-down of the restoration of fluctuations.   

\section{Conclusions}

In conclusion, we have been able to study superconducting boron-doped diamond films by using the method of amplitude-modulation of the absorbed THz radiation. By changing the frequency of the modulation we find different regimes with different values and different temperature dependences of the energy-relaxation time. The slow energy-relaxation at low temperatures is governed by a $T^{-2}$-dependence with a value of 0.7 $\mu$s at $T = 1.7$ K. At temperatures closer to $T_c$ we identify the longitudinal non-equilibrium time, in the narrow temperature range ($ 0.95 T_c < T < 0.99 T_c$). The associated inelastic-scattering time differs by an order of magnitude from the energy relaxation time found at lower temperatures. We argue that we cannot assign a conclusive interpretation to the differences in the absolute value.

Blase \emph{et al.} \cite{Blase2004} have pointed out that the superconductivity in boron-doped diamond may be intimately related to the  contribution of the stretching bond of the C atoms to the B atoms. It implies that the electron-phonon interaction leading to superconductivity is intimately related to the presence of the B atoms. The results presented here suggest that it is worth analyzing in more depth the time dependence of the nonequilibrium processes by combining the insights from Blase \emph{et al.} with insights from theories like the Sergeev-Mitin theory\cite{Sergeev2000}. 

\begin{acknowledgments}
This work was supported by the Ministry of Education and Science of the Russian Federation, Contract No. 14.B25.31.0007. The study was also implemented in the framework of the Basic Research Program at the National Research University Higher School of Economics (HSE, Russia) in 2015. T.M.K. also acknowledges the financial support from the European Research Council Advanced Grant No. 339306 (METIQUM). Alexander Semenov acknowledges the financial support of the Russian Foundation for Basic Research Grant No. 15-52-10044 and the Grant of the President of the Russian Federation No. MK-6184.2014.2 and Moscow State Pedagogical University Grant No.3.2575.2014/K. We thank A. Sergeev for helpful correspondence. 
\end{acknowledgments}

\bibliography{diamond}

\end{document}